\documentclass[pre,showpacs,twocolumn]{revtex4}
\usepackage{graphicx,amssymb,color}
\usepackage[dvipsnames]{xcolor}
\usepackage[normalem]{ulem}

\begin{document}
\title{Phase Transition in a Out-of-equilibrium Monolayer of Dipolar Vibrated Grains }

\author{Loreto Oyarte$^1$, Pablo Guti\'errez$^2$, S\'ebastien Auma\^\i tre$^2$}
\email[Corresponding author: ]{sebastien.aumaitre@cea.fr}
\author{Nicol\'as Mujica$^3$}
\affiliation{$^1$Departamento de F\'isica, Facultad de Ciencias B\'asicas, Universidad Metropolitana de Ciencias de la Educaci\'on, Avenida Jos\'e Pedro Alessandri 774, \~ Nu\~noa, Santiago, Chile \\
$^2$Service de Physique de l'Etat Condens\'e, DSM, CEA-Saclay, CNRS, 91191 Gif-sur-Yvette, France \\
$^3$Departamento de F\'isica, Facultad de Ciencias F\'isicas y Matem\'aticas, Universidad del Chile, Avenida Blanco Encalada 2008, Santiago, Chile}

\date{\today}

\begin{abstract}  
We report an experimental study on the transition between a disordered liquid-like state and a ordered solid-like one, 
in a collection  of magnetically interacting macroscopic grains.  A monolayer of magnetized particles is vibrated vertically at a  moderate density. At high excitation a disordered, liquid-like state is observed. When the driving dimensionless acceleration $\Gamma$ is quasi-statically reduced, clusters of ordered grains grow below a critical value $\Gamma_{\rm c}$. These clusters have a well defined hexagonal and compact structure. If the driving is subsequently increased, these clusters remain stable up to a higher critical value $\Gamma_{\rm l}$. Thus, the solid-liquid transition exhibits an hysteresis cycle. However, the lower onset $\Gamma_c$ is not well defined as it depends strongly on the acceleration ramp speed and also on the magnetic interaction strength.  Metastability is observed when the driving is rapidly quenched from high acceleration, $\Gamma > \Gamma_{\rm l}$, to a low final excitation $\Gamma_q$.  After this quench, solid clusters  nucleate after a time lag~$\tau_o$, either immediately ($\tau_o = 0$) or after some time lag ($\tau_o>0$) that can vary from seconds up to several hundreds of seconds. The immediate growth occurs below a particular acceleration value $\Gamma_s$ ($\lesssim \Gamma_c$). In all cases, for $t \geq \tau_o$ solid cluster's temporal growth can be phenomenologically described by a stretched exponential law. Finally, by taking into account the finite size of our system and by using simple assumptions we propose an alternative tractable theoretical model that reproduces cluster's growth, but which analytical form is not a stretched exponential~law.      
\end{abstract}

\maketitle

\section{Introduction}

Granular matter dissipates energy by friction and inelastic collisions, therefore an external driving is necessary to  
observe dynamical behaviors. Despite their intrinsic macroscopic and non-equilibrium nature, the resulting excited state, from granular gas to an ordered compact set of grains, often shares similarities with the one of  thermal systems described by  statistical physics at equilibrium. However, important non-equilibrium features emerge, like the absence of a universal effective temperature, deviations from Fourier's conduction law, spatio-temporal instabilities, the absence of scale separation, and so forth. Granular matter is therefore a good candidate to study non-equilibrium phase transitions between these different excited states \cite{lubeck,Takeuchi}.
 
The transition of granular media from liquid-like to solid-like states have been already reported. Under shearing or 
vertical tapping, monodisperse spheres tend to collectively organize themselves in the most dense crystal state 
\cite{pouliquen,lumay,zivkovic,seb}. When a monolayer of grains that is confined between two horizontal plates is 
vibrated vertically,  unexpected ordered structures nucleate from the liquid state at high excitation \cite{Prevost,Melby2005,Clerc}. 
This phase separation is driven by the negative compressibility of the effective two dimensional fluid, as in the van der Waals model for molecular fluids \cite{Clerc}. 

In the present experiment, we are concerned by granular systems where an another interaction is added to the usual 
hard sphere collisions. To do so, we use pre-magnetized spheres. Magnetized spheres have been studied under an external 
applied magnetic field \cite{paco1,paco2}. When a set of these magnetized spheres, compacted by gravity, is submitted to a vertical magnetic field, the surface of the granular packing is destabilized above a given onset and forms peaks 
\cite{paco1}. When a granular gas is submitted to an increasing horizontal magnetic field at a given excitation, a 
transition to a clustered state is observed \cite{paco2}. A mixture of magnetic and non-magnetic spheres has been also
studied in a configuration similar to ours, but at a lower volume fraction of the magnetic particles and in an horizontal 
plate with no top lid, which limits the energy injection as the system is kept in a quasi-two-dimensional state.  
In this mixed system, authors focused on the  existence and the growth in time of clusters composed of solely magnetic 
particles, in the bath of non-magnetic ones, as function of control parameters \cite{kudrolli,kudrolli2}. 

 In our study, no external field is added and the remanent magnetic moment, present in all the particles, generates a 
dipolar interaction between them. This kind of 2D dipolar liquid (with 3D magnetostatics) has been extensively studied numerically in order to  understand the transition from isolated to branched chains \cite{tlusty,duncan,tavares}. Compared to our experiments, an important difference is that these numerical studies are performed at equilibrium. Also, they use a lower surface density, $\rho = Nd^2/L^2$ up to 0.4 (were $d$ is the particle diameter, $N$ is the number of particles, and $L$ is the system size). Numerical studies also introduce the reduced temperature $T^*=k_BT/(\mu^2/d^3)$,  where $\mu$ is the magnetic moment of the sphere and $d$ their diameter. Although it is not easy to estimate it  in our experiments, 
we verified at least that the attractive force between particles with aligned moments is weak. 
Actually, two particles have to be almost in contact to 
get an attraction that overcomes their own weight, meaning that $\mu^2/d^3\sim m g d$, with  $m$ their mass and $g$  the gravity acceleration. Therefore $T^*$ can be crudely estimated by 
$T^*\sim m(A\omega)^2/(\mu^2/d^3)$, using a granular temperature proportional to the energy per grain provided by a vibration of amplitude $A$ and angular frequency $\omega$, and using  $\mu^2/d^3\sim mgd$. One gets $T^*\sim \Gamma A/d\sim 0.2-0.6$, which is actually in the range of temperature accessible numerically.  A comparison between experiments and numerical studies seems promising.   

Here, we present an experimental study of the behavior of a sub-monolayer of pre-magnetized particles 
that are confined in a shallow geometry, with a height that is small compared to the horizontal dimensions. The system is therefore quasi-two-dimensional with three-dimensional magnetostatics. The experimental cell is submitted to vertical vibration,  the control parameter being the dimensionless acceleration $\Gamma = A\omega^2/g$, which is varied keeping constant the driving frequency $\omega$. 

This paper is organized in the following way: a description of our experimental device and procedures is presented in section \ref{sec:setup}. Then, we present first, in section \ref{sec:quasi}, the behavior of the system when the control parameter $\Gamma$ is {\it quasi-statically} reduced or increased. Below a critical value of $\Gamma$, clusters organize in a hexagonal lattice. An hysteretic behavior is observed whether the acceleration of the cell is increased or decreased. In section \ref{sec:quench}, we present results starting from a liquid-like state, from which we quench the system into the hysteresis region in order to study the dynamical growth of the ordered phase.  After presenting our experimental observations on the dynamical evolution of solid clusters, we show that their growth can be fitted by a stretched exponential law. The evolution of the parameters of this law as function of the quenched acceleration is presented and the values of fitted parameters are discussed.  In section \ref{sec:model} we present a theoretical model for the solid cluster's growth, taking into account the specificity of our experiment, in particular the conserved number of particles and the constant volume. Our model reproduces well the growth but the analytical form is different to the stretched exponential law. In the last section, part \ref{sec:concl}, we present a discussion of our results as well as our conclusions. Some future perspectives are also outlined. \\
\medskip

\section{Experimental Setup and Procedures}
\label{sec:setup}

\begin{figure}[b!]
\centerline{\includegraphics[width=8cm]{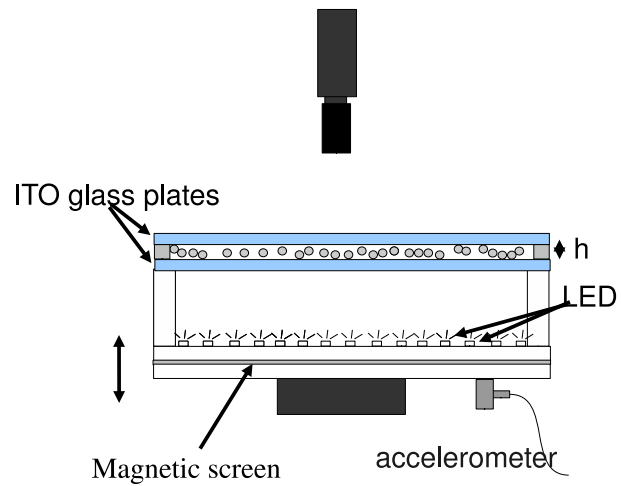}}
\caption{Sketch of the experimental setup. The cell contains the quasi-two-dimensional box where grains are confined 
between two ITO coated glass plates. The box is placed above an array of light emitting diodes and a camera pictures the particles from above. An accelerometer measures the applied dimensionless acceleration $\Gamma = A\omega^2/g$. }
\label{cell}
\end{figure}

\begin{figure*}[t!]
{\includegraphics[width=8.5cm]{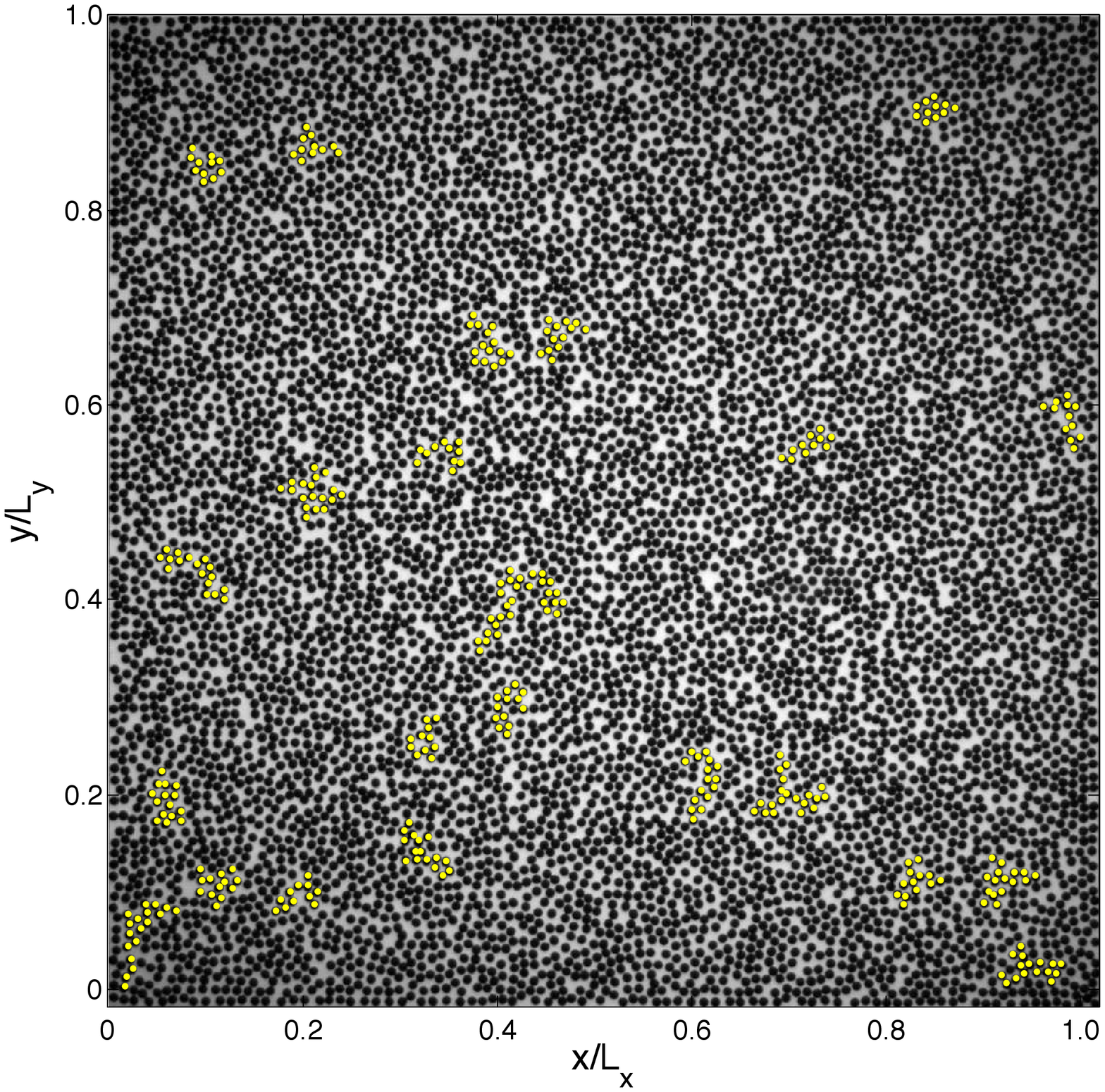}}
{\includegraphics[width=8.5cm]{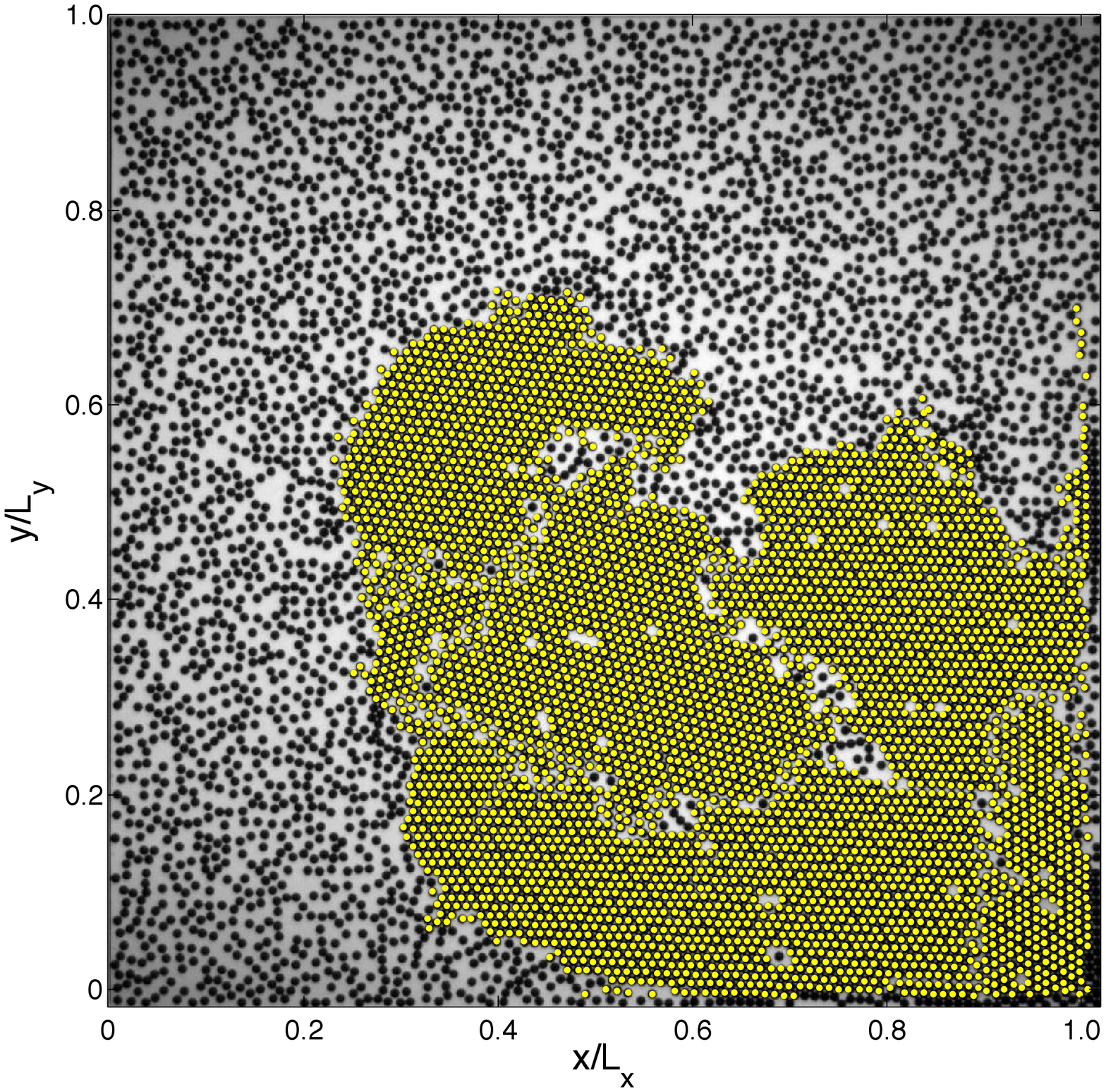}}
\caption{Left: Disordered liquid state snapshot, $\Gamma = 5$. Yellow dots show groups of more that $10$ particles that 
meet the criterium used for defining particles in a solid cluster. In this liquid state, these clusters are continuously 
forming and disappearing, tracing density fluctuations. 
Right: Coexistence of solid and liquid phases, $\Gamma = 3.5$. The solid cluster presents an hexagonal order. It is about 
two thirds of its final size. Eventually, it will be composed of about $6500$ particles, approximately $85\%$ of all 
particles. }
\label{PictCell}
\end{figure*}

The experimental device, illustrated in Fig.\ref{cell}, is similar to the one presented in references \cite{rivas,gustavo}. Stainless steel 
spheres (type 422) of diameter $d=1$ mm are confined between two horizontal glass plates, which have an indium titanium 
oxide, ITO, coating layer to prevent static electricity (resistivity $7.5\times10^{-6}$ $\Omega$m, thickness $25$ nm). The 
dimensions of the cell are: the gap between the plates $h=1.8d$, and lateral dimensions $L_x = L_y =L= 100d$. The cell is 
submitted to sinusoidal vibration $z(t) = A \sin(\omega t)$, provided by an electromagnetic shaker. We take care to screen 
its magnetic field. The cell is illuminated from bellow by an array light emitting diodes. A camera takes pictures from 
the top. Typical images are shown in Fig. \ref{PictCell}. Almost all particles are tracked using an open source Matlab 
code \cite{matlabcode}, which works well for the sub-monolayer surface filling fractions that are used. The cell acceleration, 
$a(t) = \ddot z (t)$, is  measured with a piezoelectric accelerometer fixed to the base. Our control parameter 
is the dimensionless acceleration $\Gamma = A\omega^2/g$. Results presented here have been obtained with filling fraction 
density  $\phi=N\pi d^2/4 L_xL_y = 0.59$ ($\rho = 0.75$), where $N=7500$ is the number of spheres, frequency 
$f=\omega/2\pi = 100$ Hz, period of base oscillation $P= 1/f = 0.01$ s, and $\Gamma = 3 - 5.3$. We have verified 
that the phenomenology is not qualitatively affected by the number of particles as long as the system is not too dense. 

A major experimental challenge is to keep constant the magnetization of the spheres. Indeed, depending on the 
magnetization procedure, some shifts occur for the acceleration onset values, although the qualitative 
behavior of the system remains the same. Grains are magnetized by contact with a strong Neodyme magnet. Some additional thermal treatments have also been probed to maintain the magnetization. The magnetic moment acquired by the spheres is partially lost when the system is vibrated for long times. This loss of interaction strength is produced by collisions and local particle heating.  
This issue will be discussed in more detail below, in particular when we present experimental results of quasi-static acceleration ramps.

Starting from an homogeneous liquid state at $\Gamma = 5$, and decreasing $A$ for $\omega$ fixed, we observe that for $\Gamma \le 3.5$ clusters of ordered grains grow almost immediately, as shown in Fig. \ref{PictCell}-right. In contrast to numerical simulation predictions where all particles form chains \cite{tlusty,duncan,tavares}, 
here clustered particles are well organized in a two dimensional triangular crystal, except at the edge of the cluster 
where more linear chains coexist with disordered liquid-like particles. In fact, the precise cluster's topology depends on $\Gamma$ and the speed at which the solid cluster grows. For rapid changes in driving  from high acceleration to $\Gamma \sim 1-3$, the solid phase will grow quickly and will be first formed by many sub-clusters separated by defects as well as 
many linear chains. However, if driving is varied slowly from a liquid state, a single cluster will first nucleate 
and then grow slowly, with a smaller amount of defects, like the one shown in Fig. \ref{PictCell}-right.

In order to define the number of particles that are in a solid cluster, $N_c$, we compute the area of the Delaunay 
triangle, $\mathcal A$, {\it i.e} the area of all the triangles joining the center of each particle with two of this nearest 
neighbors. All particles belonging into a Delaunay triangle for which $\mathcal A< 0.5d^2$ are defined as belonging to 
a cluster.  The precise value of the onset does not modify qualitatively the results presented here. It has been chosen in 
order to find all the particles inside a cluster in the cell, as the example shown in Fig. \ref{PictCell}-right. Some particles 
in the liquid phase will inevitably fit this condition as well. This will add a background noise about $1000$ particles 
belonging to small unstable clusters, tracing density fluctuations, like those shown in Fig. \ref{PictCell}-left.

\section{Experimental Results}
\subsection{Quasi-static acceleration ramps}
\label{sec:quasi}

\begin{figure}[t!]
\centerline{\includegraphics[width=8.5cm]{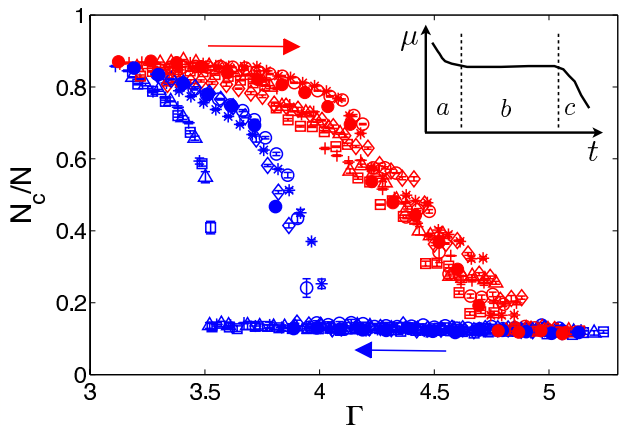}}
\caption{Total numbers of particles inside a cluster $N_c/N$ as function of $\Gamma$. Acceleration is varied 
{\it quasi-statically}. Blue (red) symbols indicate the results when $\Gamma$ is reduced (increased). Several ramps are shown,
for different realizations. Each pair of symbols correspond to a closed $\Gamma$ loop, realized in this order: 
$\color{blue} \circ$, $\color{red} \circ$, $\color{blue} *$, $\color{red} *$, $\color{red} \diamond$, $\color{blue} 
\diamond$, $\color{blue} +$, $\color{red} +$, $\color{blue} \triangle$, $\color{red} \triangle$, $\color{blue} \square$, 
$\color{red} \square$, $\color{blue} \bullet$, $\color{red} \bullet$. For each ramp, $\Gamma$ is varied with a waiting time 
of $15$ s before obtaining images during $5$ s at $1$ fps. There are two exceptions: ramps for $30$ s ($\color{blue} +$, 
$\color{red} +$) and $180$ s ($\color{blue} \bullet$, $\color{red} \bullet$) waiting times. Magnetization is reduced after 
many hours of experiments, evident in some of the the decreasing ramps ($\color{blue} +$, $\color{blue} \triangle$, 
$\color{blue} \square$), were the transition $\Gamma$ is reduced, although not so in their correspondent increasing ramps 
($\color{red} +$, $\color{red} \triangle$, $\color{red} \square$). The last two ramps correspond to a longer waiting 
time ($180$ s, $\color{blue} \bullet$, $\color{red} \bullet$) for which the transition point again is practically the same 
as before. Inset: schematic representation of magnetic moment dependence on time.}
\label{Adiagrowth}
\end{figure}

We first present results obtained performing {\it quasi-static} cooling and heating ramps, by slowly decreasing and increasing the driving acceleration respectively. In~Fig.~\ref{Adiagrowth}, we plot the fraction of particles inside a cluster $N_c/N$ 
as function of $\Gamma$. Blue symbols represent the fraction of crystallized particles when the acceleration is slowly 
decreased from the liquid state, whereas red ones represent this quantity when the acceleration is slowly increased. 
Each similar pair of symbols correspond to different experimental realizations. 
More precisely, the procedure is the following: first the system is set at high 
acceleration ($\Gamma \approx 5$). The system is left to evolve to a stationary state during a waiting time, $t_w$. Then, 
five images are acquired at a rate of $1$ fps. Immediately after, $\Gamma$ is reduced by a small amount and the system is 
again left to evolve during $t_w$ until the next acquisition of images. This procedure is repeated until we reach the 
lowest acceleration of about $\Gamma=3.1$. Then, the same procedure is used but increasing $\Gamma$ until the completely 
fluidized state is obtained again. For most of the symbols in Fig. \ref{Adiagrowth}, $t_w=15$ s (i.e $t_w = 1500P$). 
For one case this loop has been realized in the opposite way (symbols $\color{red} \diamond$, $\color{blue} \diamond$ in 
Fig. \ref{Adiagrowth}), with no difference in the resulting curves. In two cases, the waiting times are different: ramps 
with waiting times of $30$ s ($t_w= 3000P:$ $\color{blue} +$, $\color{red} +$) and $180$ s ($t_w= 18000P:$ $\color{blue} 
\bullet$, $\color{red} \bullet$). In the caption of Fig. \ref{Adiagrowth}, we indicate the order in which these 
ramps were realized. Particles were not re-magnetized during all the experiments presented in this figure.

The inset of Fig. \ref{Adiagrowth} shows an schematic representation of the variation of the average magnetic dipole moment $\mu$ with time. Here, ``time" means vibration oscillations, {\it i.e.} time during which the particles are in a fluidized collisional state at a given $\Gamma$. Three stages are identified ($a$, $b$ and $c$), and the vertical dotted lines indicate the transitions between them. The exact positions of the different transitions depend on $\Gamma$ and the particular magnetization procedure. Stage $a$ is characterized by a fast decrease of $\mu$; for $\Gamma \sim 5$, this can endure for a few tenths of minutes. Later, stage $b$ corresponds to a stable average magnetization, which typically lasts for several hours. And finally, after many particle collisions, $\mu$ continues to decay. 

The main result of Fig. \ref{Adiagrowth} is that, for the current cooling and heating rates and procedure, there is a stable loop with two transition points: from an homogeneous liquid to 
solid-liquid coexistence at $\Gamma_c = 3.9\pm 0.1$, and from solid-liquid coexistence to the homogeneous liquid state at 
$\Gamma_l = 4.8 \pm 0.1$. Thus, when the driving is slowly increased, the last cluster disappears always at a critical 
value which is higher than the value at which the first crystal appears when the driving is decreased: $\Gamma_l > \Gamma_c$. 
We will refer to this loop as the {\it quasi-static hysteresis loop} (stage $b$), although it is not reproducible 
for very long experiments because of a reduction in particle's magnetization (stage $c$).

Indeed, some curves obtained for 
decreasing ramps  have different $\Gamma_c$ values as shown by the symbols $\color{blue} +$, $\color{blue} \triangle$,  and $\color{blue} \square$ in Fig. \ref{Adiagrowth}. These shifted curves, which result in a lower transition acceleration 
value ($\Gamma_c \approx 3.5$) from the liquid state to the coexistence of solid and liquid phases, correspond to later runs without remagnetization in between. Therefore, these shifts are mainly due to particle demagnetization. However, for such runs, $\Gamma_l$ does not change. We remark that when the waiting time $t_w$ is increased to 180 s ($18000P$), symbols $\color{blue} \bullet$, $\color{red} \bullet$ in Fig. \ref{Adiagrowth}, even with a lower magnetization the previous hysteretic cycle is recovered. It seems that for a lower magnetization and with $t_w=1500P$ or $t_w=3000P$, we do not wait enough the obtain the crystal growth. Thus, the loss of magnetization seems to increase the time necessary to get a quasi-stationary state. However, beyond this variability, the main qualitative feature remains: the  hysteretic behavior with $\Gamma_c < \Gamma_l$. 

Finally, measurements not shown here that are performed just after remagnetization (stage $a$) show that both $\Gamma_c$ and $\Gamma_l$ are shifted to higher values, whereas measurements made later, for very long experimental times after magnetization (very late times in stage $c$), present both $\Gamma_c$ and $\Gamma_l$ shifted to lower values.

\begin{figure}[t!]
\centerline{\includegraphics[width=8.5cm]{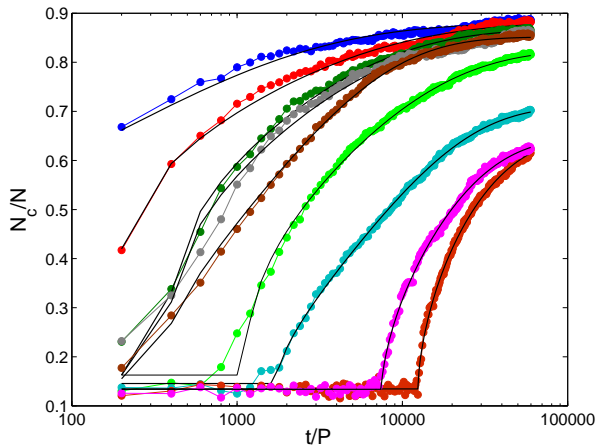}}
\caption{Number of particles that form part of a solid cluster, $N_c/N$, versus  time $t/P$, for several  $\Gamma_q$ (quenched state). Initial time $t=0$ corresponds to the moment $\Gamma$ is abruptly quenched from $\Gamma = 5$ to $\Gamma_q$. For each run, images are acquired during $10$ min, equivalent to $6\times10^4 P$. From left to right $\Gamma_q= 3.15$, $3.31$, $3.42$, $3.53$, $3.63$, $3.74$, $3.94$, $4.13$, and $4.02$. For $\Gamma_q = 3.80$ and $4.2$ there was no transition within the observation time. Continuous lines show fits for stretched exponential growth law, Eqn. (\ref{strexp}). In average, for increasing $\Gamma_q$, $\tau$ and $\tau_o$ increase and $N_\infty$ decreases.}
\label{tpstrace}
\end{figure}

\subsection{Quench experiments}
\label{sec:quench}

Knowing  the onset of crystallization, we now study the dynamics when the homogeneous liquid phase is quenched into a 
final state near the solid cluster nucleation. Starting from  $\Gamma=5$, we reduce suddenly the driving to a 
final value $\Gamma_q$ between $3$ and $4.5$. We then acquire images at a low frame rate ($0.5$ fps) for several minutes. 
The typical measurement time is $10$ min ($t_m=6\times 10^4P$) in order to keep constant the particle's magnetic dipole moment during complete run (that is for all the quenches), although some runs with $t_m = 20$ min have also been realized ($t_m = 1.2\times 10^5P$). 
For each image, we compute $N_c$, the total number of particles that belong to a solid cluster. 

The temporal evolution of $N_c$ is shown in Fig. \ref{tpstrace} for different 
values of $\Gamma_q$. This series of quench experiments was performed in between the quasi-static loops identified by symbols $\color{blue} \diamond$ and $\color{blue} +$ in Fig. \ref{Adiagrowth}. Two time scales are clearly present: a growth time $\tau$ and a time lag $\tau_o$. Clusters grow 
slower for higher $\Gamma_q$. For $\Gamma_q \lesssim 3.5$, clusters grow immediately ($\tau_o = 0$); for higher 
$\Gamma_q$, cluster's nucleation is delayed ($\tau_o>0$). This allows us to define for this realization $\Gamma_s \approx 3.5$ as the acceleration below which $\tau_o \approx 0$. In the case $\tau_o>0$ the system is metastable: it can be either in an homogeneous liquid phase or in a solid-liquid phase separated state.
The transition from the former to the latter occurs if there is a density fluctuation strong enough to nucleate crystallization. Additionally, this density fluctuation has to have particles aligned in such a way that they can bound together. As $\Gamma_q$ approaches $\Gamma_l$, this fluctuation has to be stronger (more dense), because the granular temperature is higher. 
Thus, it becomes less probable too. However, at the same time, its final size is smaller, requesting less bounding energy. 
The corresponding lag time $\tau_o$ becomes usually larger, but its dependence on the appearance of the correct density fluctuation makes it highly variable from one realization to another. The general results is valid for every quench experiment done in the {\it stable hysteresis loop}:  For $\Gamma_q$ approaching $\Gamma_l$, the crystal growth time $\tau$ increases, the asymptotic number of particles in the solid phase decreases, and the lag time $\tau_o$ seems to increase in average but its variance too (much more independent realizations are needed to study $\tau_o$ properly).

\begin{figure*}
\includegraphics[width=8.5cm]{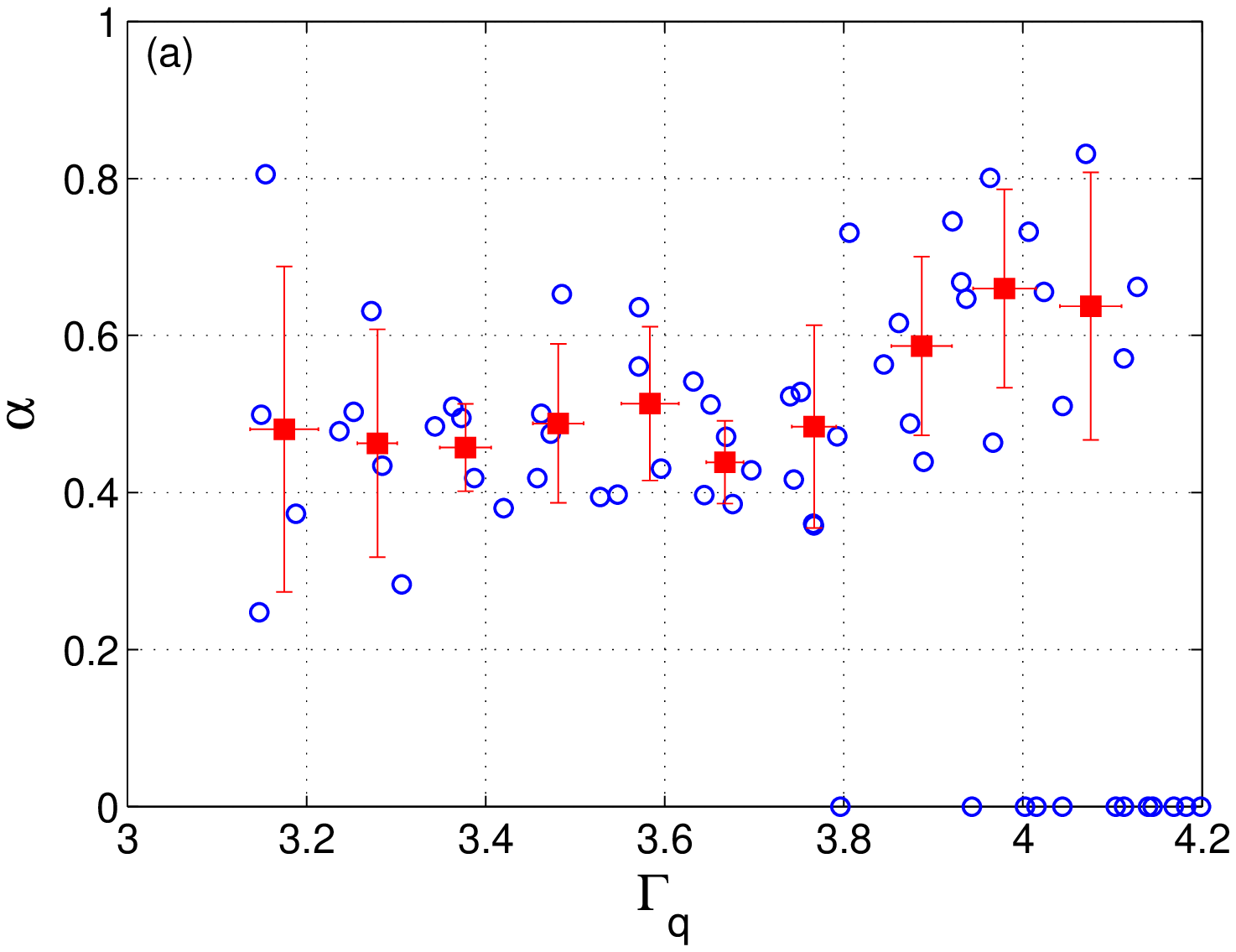}
\includegraphics[width=8.5cm]{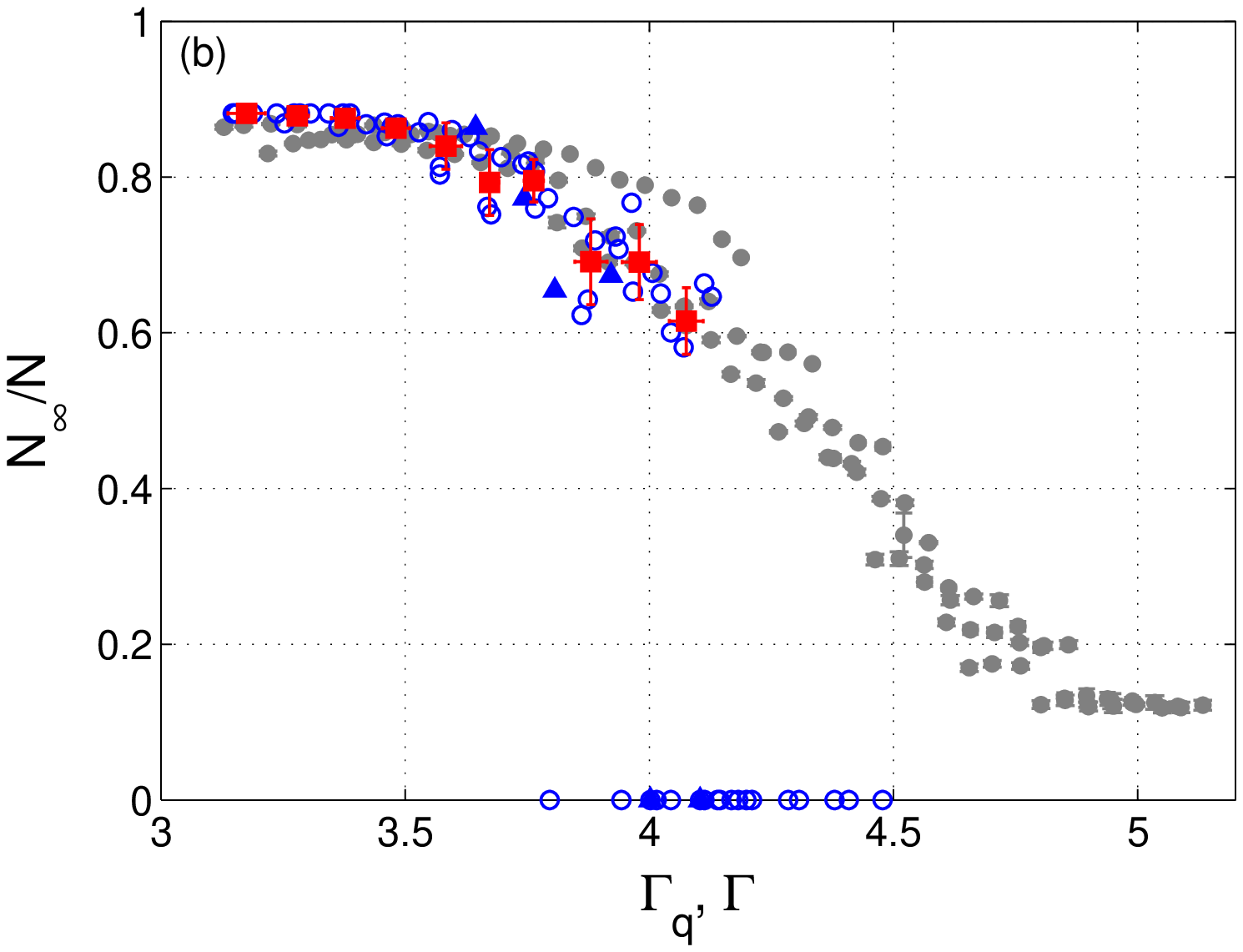}
\includegraphics[width=8.5cm]{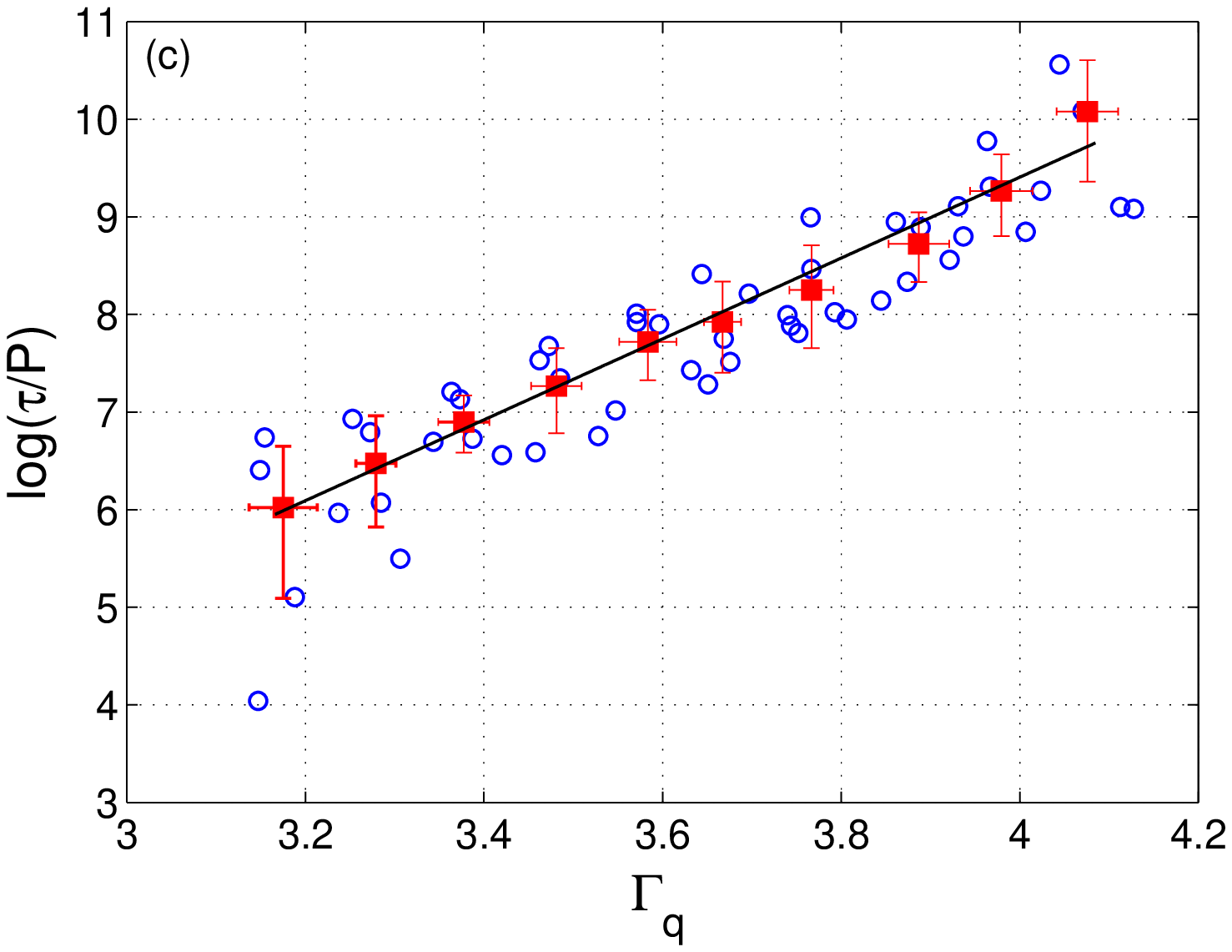}
\includegraphics[width=8.5cm]{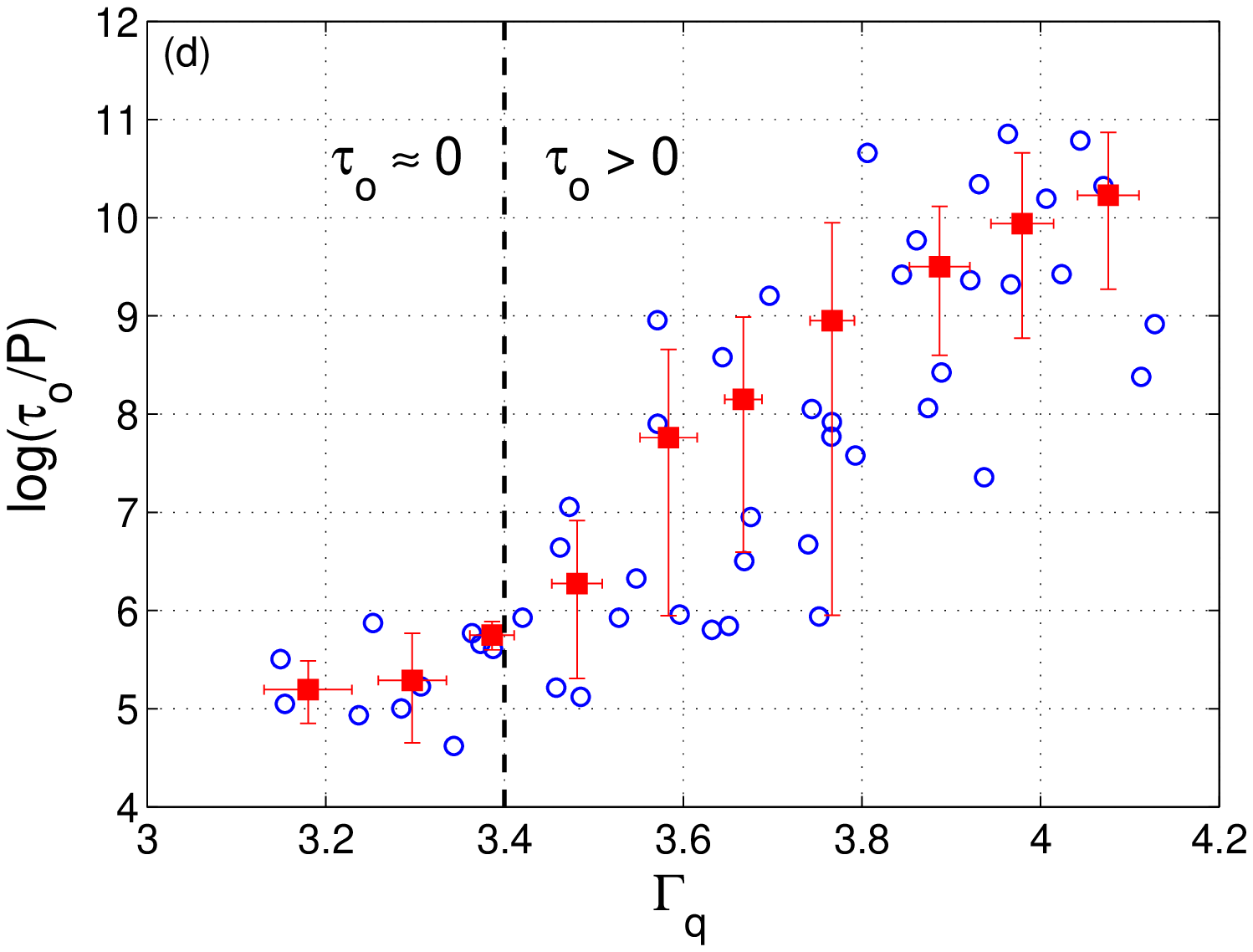}
\caption{(Color online) Fitted parameters $\alpha$ (a), $N_\infty$ (b), $\tau$ (c) and $\tau_o$ (d) as functions of $\Gamma_q$. For (a) and (b) symbols on the x-axis correspond to realizations for which the phase separation was not achieved on the experimental observation time. Open circles ($\color{blue} \circ$) represent raw data obtained from several independent ramps, whereas solid squares ($\color{red} \blacksquare$) correspond to averages of data for subsets of $\Gamma_q$. Errorbars are computed from standard deviations. In (b), open circles ($\color{blue} \circ$) have a total measurement time $t_m = 6\times10^4 P$, but some data is shown for $t_m = 12\times10^4 P$ ($\color{blue} \blacktriangle$). Solid circles ($\color{gray} \bullet$) show some of the quasi-static heating curves from Fig. \ref{Adiagrowth}. In (c) and (d) the growth time $\tau$ and lag time $\tau_o$ are plotted in semi-log scale. Both times increase strongly as $\Gamma_l = 4.8$ is approached. The continuous line in (c) shows an linear fit $\log(\tau/P) = a^*\Gamma_q+\tilde a$, with $a^* = 4.1\pm0.5$. The dashed line in (d) indicates the boundary below which clusters grow immediately after a quench.}
\label{linreg}
\end{figure*}

In Fig. \ref{tpstrace} we also present our data fitted by a stretched exponential  growth law
\begin{equation} 
N_c(t) = N_\infty + (N_o - N_\infty)\exp[-[(t-\tau_o)/\tau]^\alpha],
\label{strexp}
\end{equation}
where $N_\infty$ is the asymptotic number of particles in the solid phase and $\tau$ the growth time. $N_o$ is the background ``noise" in the measurement of the number of solid particles; in the liquid state, there are many small short-living clusters that are considered as composed by particles in a solid phase, as shown in Fig. \ref{PictCell}. Together with the exponent $\alpha$ and time lag $\tau_o$ these are used as fit parameters. This stretched exponential law has been used in first order phase transitions to model the growth of the stable phase into the metastable one \cite{avramiI}. It has also been used to describe the compaction of a sand pile under tapping~\cite{lumay}. In our case, the adjustment shown by the continuous lines in Fig. \ref{tpstrace} is very good. The 
exponent $\alpha$ fluctuates around $1/2$ as function of $\Gamma_q$, as shown in~Fig \ref{linreg}a (data on the x-axis correspond to realizations for which the phase separation was not achieved on the experimental observation time). In fact, the results 
that are discussed in what follows do not depend strongly on the fact that $\alpha$ can be left as a free parameter or 
fixed to $\alpha = 0.5$. 

The dependence of $N_\infty$ on $\Gamma_q$ is shown in Fig. \ref{linreg}b.  
The asymptotic fraction of particles in the solid phase, $N_\infty/N_c$, decreases for high quenching accelerations, 
from $\approx 0.9$ for $\Gamma_q=3-3.4$ to $\approx 0.6$ at $\Gamma_q\approx4$.  We notice that except for some experiments where we do not wait enough to overcome the time lag $\tau_o$ 
(data on the x-axis),  the value of $N_\infty(\Gamma_q)$ collapses with the one obtained for $N_c(\Gamma)$ by {quasi-static} heating, showing that we really perform an adiabatic modification of the forcing for the heating case with our quasi-static procedure. 

In Figs. \ref{linreg}c and \ref{linreg}d we present $\tau$ and $\tau_o$ as functions of $\Gamma_q$.  Our results show that the both times increase strongly in a small range of $\Gamma_q$: about two orders of magnitude for $\Gamma_q$ between $3.1$ and $4.1$. The longest measured growth and lag times,  for $\Gamma_q \approx 4.1$, are $\tau \approx 3.5\times10^4 P = 350$ s and $\tau_o \approx 5\times10^4 P = 500$ s. 
If $\Gamma_q$ is approached to $\Gamma_l = 4.8$, the lag time $\tau_o$ becomes larger than the measurement time $t_m$ and no transition from the homogeneous fluid state to the coexistence between solid and fluid phases is observed. In the short  parameter range where equation (\ref{strexp}) can be verified, 
we obtain $\tau \sim \exp({a^*\Gamma_q})$, with $a^* = 4.1\pm0.5$. Concerning the time lag $\tau_o$, we observe that it is highly variable; much more experiments seem necessary in order to obtain significant statistics. For $\Gamma_q < \Gamma_s = 3.4$, this time lag tends to a constant, $\tau_o \sim 1$ s. As we acquire images at 0.5 fps, this saturation seems artificial. In fact, after doing many realizations we conclude that for  $\Gamma_q <\Gamma_s = 3.4$ the solid cluster seems to grow immediately with no delay. Notice that defined this way, $\Gamma_s$ is different and lower that $\Gamma_c$. Thus, our current measurements do not allow us to measure small values of $\tau_o$ with enough precision. However, for   $\Gamma_q > \Gamma_s = 3.4$, $\tau_o$ is a few tenths of seconds, and, despite the poorer statistics,  a strong growth is observed. Finally, the strong increases for both $\tau$ and $\tau_o$ is the reason we can not get closer to $\Gamma_l$ within the experimental observation times. For example, for $\Gamma_q = 4.5$ and $\Gamma_q = 4.8$, the extrapolation for $\tau$ gives $\tau \sim 10^5 P \sim 1000$ s and $\tau \sim 4\times10^5 P \sim 4000$ s respectively. Because $\tau_o$ also would increase significantly, then very long measurement times will be needed and particles would decrease significantly their magnetic interaction strength during these experiments.

\section{Detailed balance model for crystal growth}
\label{sec:model}

In order to model crystal growth in our system, we need to consider a few important considerations: first, the transition occurs at constant volume, which in practice implies a constant surface $L^2=S_c(t)+S_l(t)$, with $S_c(t)$ the surface occupied by the (2D) solid cluster at time $t$ and $S_l(t)$ the surface available for the disordered liquid phase at that time. Also, the number of particles is conserved, $N=N_c(t)+N_l(t)$, where $N_l(t)$ is the number particles in the the disordered phase. Secondly, we will consider that clusters grow with a circular shape. Because of the well defined structure of the cluster, once $N_c(t)$ is determined, then $S_c(t)$ is known too. Indeed, they are just related by a geometrical factor: $S_c=\pi R(t)^2$, where $R(t) =b\sqrt{N_c(t)}$ is the mean cluster radius at time $t$ and $b$ a geometrical constant of order of the grains diameter. For instance, for a perfectly ordered crystal in hexagonal close packing one gets
\[
\frac{b}{d}=\left(\frac{\sqrt{3 }}{2\pi}\right)^{1/2}=0.53.
\] 
Next, we can establish a {\it detailed balance} equation to determine the time evolution of $N_c (t)$. This can be written
\begin{equation}
\frac{dN_c(t)}{dt}=p_{l\rightarrow c}N_l(t)-p_{c\rightarrow l}N_c(t),
\label{DetBal}
\end{equation}   
\noindent
where $p_{l\rightarrow c}$ is the probability by unit  time that a particle of the disordered phase is captured by the solid cluster and $p_{c\rightarrow l}$ is the probability by unit time that a particle escapes from the cluster to the liquid phase. The first can be estimated crudely as the collision frequency between a cluster and the liquid particles which have a velocity small enough to be caught  by the former. If we characterize these particles by their mean velocity $v_o$, we assume  
\[p_{l\rightarrow c}\sim\frac{N_l(t)}{S_l(t)}v_o 2 R(t)
\] 
since $2 R(t)$ is almost the cross-section of the cluster (for $R(t)>>d)$. Now, particles leaving a cluster have to be at its perimeter. We can assume that they have a constant  probability by unit  time $p$ to leave this perimeter. Therefore, because for simplicity we consider an almost circular cluster we obtain 
\[
p_{c\rightarrow l}\sim\frac{2\pi R(t) b}{\pi R(t)^2} p.
\] 
Both $v_o$ and $p$ would depend on quench acceleration $\Gamma_q$, on the distance to $\Gamma_l$, that is $\Delta \Gamma=\Gamma_l-\Gamma_q$, on the particle density $N/L^2$ and on dipole magnetization $\mu$. Knowing the relation between $L^2$, $S_l(t)$, $R(t)$, $ N$, $N_l(t)$ and $N_c(t)$ imposed by the constant surface, number of particles and the crystal structure of the clusters, we can rewrite (\ref{DetBal})~as
 \begin{equation}
\frac{dn(\theta)}{d\theta}=\left(\frac{\left(1-n(\theta)\right)^2}{1-\beta n(\theta)}-\gamma\right)\sqrt{n(\theta)},
\label{DetBalII}
\end{equation}    
where we haved introduced the normalized number of particles in the solid cluster, $n(\theta)=N_c(t)/N$, the rescaled time, $\theta=(2\rho v_o b \sqrt{N_o})t$, with $\rho=N/L^2$, the geometrical factor $\beta=\pi b^2\rho$ and the parameter $\gamma={p}/({b\rho v_o N})$. Notice that these parameters are not free fitting parameters. Indeed, the geometric parameter should be $\beta\approx\rho d^2\sqrt{3}/2\approx 0.65$. It will be fixed to this value hereafter. Moreover,  in the time-asymptotic regime or along the quasi-static heating ramp we can impose ${dn(\theta)}/{d\theta} = 0$. Then, the parameter $\gamma$ is obtained through the the well defined stationary values $n_\infty = N_\infty/N$ shown Fig. \ref{linreg}b. Therefore, for a given value of the quenched acceleration $\Gamma_q$, the value of $\gamma$ is obtained by 
\begin{equation}
\gamma=\frac{\left(1-n_\infty\right)^2}{1-\beta n_\infty}.
\label{gammavsninf}
\end{equation}
The only undetermined parameter is the characteristic velocity $v_o$,
thus the time scale $t^*=(2\rho v_o b \sqrt{N})^{-1}$.

\begin{figure}
\centerline{\includegraphics[width=8.5cm]{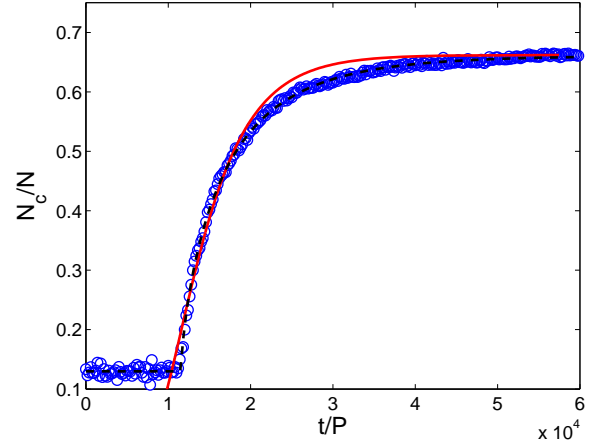}}
\caption{(color online) The adjustment of experimental results with equation (\ref{tauvsn}) is shown with a continuous curve (red) for $\Gamma_q=3.92$ for which $N_\infty/N=0.662$ and $\beta=0.65$. Fit parameters are the time scale $t^*/P=4000$ and growth starting time $t_o/P=6500$. For comparison the dashed line (black) shows the stretched exponential fit given by Eqn. (\ref{strexp}), with parameters $N_o /N= 0.130\pm0.001$, $N_{\infty}/N = 0.661\pm0.001$, $\alpha = 0.77\pm0.02$, $\tau/P = 5350\pm80$, and $\tau_o/P = 11520\pm50$. }
\label{fittau}
\end{figure}

Equation  (\ref{DetBalII}) can be solved to obtain the following expression for $\theta$ as a function of $n$:   
 \begin{eqnarray}
\frac{\theta}{\theta_ o} &= C_1\cdot \tanh^{-1}\left(\sqrt{\frac{n}{n_\infty}}\right)-C_2\cdot \tanh^{-1}\left(\sqrt{\frac{n}{n^*}}\right)\\
&=\ln\left[\left(\frac{1+\sqrt{n/n_\infty}}{1-\sqrt{n/n_\infty}}\right)^{C_1/2} \cdot\left(\frac{1-\sqrt{n/n^*}}{1+\sqrt{n/n^*}}\right)^{C_2/2}\right]
\label{tauvsn}
\end{eqnarray}  
\noindent
where 
\[
n^*= \frac{(2-n_\infty-\beta)}{(1-\beta n_\infty)}
\]
is the second root (larger than 1) of the polynomial relation issued of (\ref{gammavsninf}),  
\[
C_1= \frac{(1-\beta  n_\infty)}{\sqrt{n_\infty}},
\]
\[
C_2= \frac{(1-\beta)^2}{\sqrt{(2-n_\infty-\beta)(1-n_\infty\beta)}},
\] 
and
\[
\theta_o= \frac{2(1-\beta n_\infty)}{(2-n_\infty-\beta)(1-n_\infty\beta)}.
\]
Fig. \ref{fittau} shows that a numerically inverted relation (\ref{tauvsn}) with $t^*$ as a fit parameter reproduces also quite satisfactorily the experimental results. We stress that this fit has only two adjustable parameters, $t^*$ and $t_o$ (the starting growth time), whereas the stretched exponential form has five, $N_o$, $N_\infty$, $\tau$, $\tau_o$ and $\alpha$ (when $\alpha$ is left as a free parameter). 

\section{Discussion and Conclusions}
\label{sec:concl}

In summary,  quasi-static cooling and heating of a layer of dipolar interacting spheres is performed by slow changes of the driving acceleration.
The system exhibits an hysteretic phase
transition between a disordered liquid--like phase and an ordered phase. The ordered phase is formed by dense clusters, structured in
an hexagonal lattice,  coexisting with disordered grains in a liquid-like phase. By quenching the system in the vicinity of the transition, we study the dynamical
growth of the ordered clusters, which can arise after a time delay. The time evolution of the number of particles in the solid-crystal phase is well fitted by a stretched exponential law, but also by relation~(\ref{tauvsn}) that we have obtained analytically.  

The first point to discuss is why we do not observe particles in a state of branched chain, as reported in numerical 
simulations \cite{tlusty,duncan,tavares}.
Indeed our filling fraction, $\phi = 0.59$ as well as the reduced temperature $T^*\sim 0.4$, are not so far from the values used in \cite{tlusty,duncan,tavares}. 
However, we do not observe the branched state reported in these studies, at least not as a stable, stationary configuration. For our experiment, as mentioned in section \ref{sec:setup}, if a quench is done deeply into the solid phase, $\Gamma_q \sim 1 - 3$, then many small solid clusters are formed quickly and many linear chains are present. However, this state eventually evolves to a more densely packed state, with almost no linear chains, in coexistence with the liquid phase. More systematic experiments should be performed to study properly this aging process.  

Compared to numerical simulations, one of the major differences is in the vertical vibration imposed in our experiment to sustain particle motion. Indeed, instead of the 2D thermal motion used in numerical studies, a particle in the solid phase collides the bottom
and top plates during an excitation period. The vertical motion might disadvantage the formation of long chains, and compact clusters
might be more stable under such driving. Moreover, in contrast to numerical simulations, the magnetic dipole moments are probably not uniformly distributed in all the particles. Although it is really difficult to estimate the width of dipole moments distribution in our experiment, a faction of them could be small enough to prevent bounding.

The second point deserving discussion is the existence of the quasi-static hysteresis loop when the driving amplitude is slowly modified. A crucial point is that during a cooling ramp there is a specific driving, $\Gamma_c$, which depending on the particular magnetization history is between $3.5$ and $3.9$, below which solid clusters grow. Due to magnetization reduction and because $\tau$ and $\tau_o$ vary with driving amplitude, this specific boundary is very difficult to determine more precisely. How slow one should vary the driving depends on the time scales that are present in the system, that is on $\tau$ and $\tau_o$. The quench experiments show that, within the hysteresis loop, both times grow strongly with driving amplitude. These same quench experiments show the existence of another particular acceleration $\Gamma_s$ ($\lesssim \Gamma_c$), below which $\tau_o \approx 0$. This onset value seems to be better defined that $\Gamma_c$. 

Our quasi-static increasing acceleration ramps show that the heating branch of the loop seems to be a stable, adiabatic branch. In particular, all quasi-static heating ramps follow the same curve and the critical driving above which solid clusters disappear, $\Gamma_l = 4.8\pm0.1$, is very reproducible. The fact that we are able to follow adiabatically this stable branch can be understood with the following reasoning: 
\begin{itemize}
\item A ramp starts at low $\Gamma$ with a well formed solid cluster (or clusters) in coexistence with the liquid phase; 
\item The driving amplitude is increased in a small amount, which increases the system's kinetic energy (granular ``temperature''); 
\item Particles in the solid phase but at the cluster's boundary will receive stronger collisions from those in the liquid phase, so the probability of getting ripped off increases; 
\item This results in a cluster size reduction, more or less continuously as the driving is increased further;
\item As the change from the one state to another occurs smoothly, the time it takes can be short.
\end{itemize} 
The fact that $\Gamma_l>\Gamma_s$ demonstrates that there is a metastable region for which the system might stay in a completely fluidized state, but from which eventually a solid cluster will nucleate and grow until a stationary size is reached. Also, quench experiments done for $\Gamma_q = 3.4 - 3.9$ show that the final cluster size is the same as the obtained for the heating ramp ($N_\infty/N$ collapses with $N_c/N$ of the heating quasi-static loop). This shows that the cooling ramp branch of the quasi-static loop is not stable, implying that the cooling ramp was probably not done slowly enough. This is consistent with the fact that for a given magnetic interaction strength, the value of $\Gamma_c$ depends on the decreasing ramp rate. As the number of solid particles has to grow from the background noise level to the final asymptotic one, then waiting times much longer that $\tau_o$ are necessary. Another possible route for the metastable state characterization would be to study precursors through density and velocity correlations.

We now turn to the values of $\alpha$ used to fit experimental data with a stretched exponential law. At equilibrium, we would expect an exponent $\alpha=D+1=3$ \cite{avramiI}, where $D=2$ is the spatial dimension, but instead we obtain $\alpha \approx 0.5$.  Due to the anisotropy of the attractive force, 
of the special shape of the interface and of the fact that most of the clusters start to grow from boundaries, one could assume that our experiment behaves more like 1D system, but even in 
this extreme case, we should have get $D\geq1$ and $\alpha\geq2$. Moreover, one  has to underline that we are in a out-of-equilibrium dissipative system. Others aspects can therefore play a role, like dissipation, which is probably different in the cluster and liquid phase, or the very particular thermalization imposed by vertical vibration.
This effect could be especially important near the transition where a critical behavior can be expected. Actually, an exponent $\alpha<1$  is measured in another out-of-equilibrium system:  the slow
compaction of granular packing by tapping \cite{lumay}, but the meaning of such law is still under debate in this case.  This issue deserves further studies by changing  the boundary shape (from square to circular) or by doing the experiment in a viscous fluid to modify the dissipation or by exciting the grains by vibration using colored noise with a broad frequency range.

It is clear that the present system is strongly out of equilibrium. Despite it can reach stationary states, at coexistence the effective in plane ``granular" kinetic temperatures can be very different between the liquid and solid phases. In the metastable region, the transition from the liquid state to the coexisting one can be achieved with the experimental time scales that are available. This transition occurs because the appropriate fluctuation occurs in the liquid state. But, once a particle is trapped in the solid state, specially those deep inside the boundary layer, it is very difficult for them to escape. Indeed, from our model we can estimate in a stationary regime that the ratio between the probability rates $p_{l\rightarrow c}$ and $p_{c\rightarrow l}$ is given by 
\begin{equation}
\frac{p_{l\rightarrow c}}{p_{c\rightarrow l}} = \frac{n_\infty}{1-n_\infty}.
\end{equation}
For $\Gamma<\Gamma_s$, $p_{l\rightarrow c}/p_{c\rightarrow l} \gtrsim 7$. This allows us to propose that the transition in our system is a good candidate to be considered as quasi-absorbing hysteretic phase transition. However, the relative small size of the system and our imperfect control on the spheres magnetization and therefore on the transition onsets, makes very difficult any
quantitative benchmark between our system with other systems exhibiting out-of-equilibrium quasi-absorbing phase transitions. Nonetheless, some promising qualitative similarities can be underlined between our experiment and the out-of-equilibrium transition observed between two turbulent phases of a liquid crystals, transition that belongs to the direct percolation universality class \cite{Takeuchi}. In this system an hysteresis loop is observed between a turbulent phase stable at high excitation (called DSM2) and a quasi-absorbing state (called DSM1) \cite{kai,takeuchi2}. Also, an exponential decay of DSM2 phase is observed when it is quenched in the DSM1 stable region \cite{Takeuchi}. Computations of the critical exponents is beyond the experimental accuracy accessible with our system and procedures and will need a larger device and also a better control of the magnetic interaction strength.

\acknowledgments

We thank Cecile Gasquet, Vincent Padilla and Correntin Coulais for valuable technical help and discussions. This research was supported by grants ECOS-CONICYT C07E07, Fondecyt  No. 1120211, Anillo ACT 127, and RTRA-CoMiGs2D. 
 

\end{document}